# A European Multi-Center Breast Cancer MRI Dataset

Gustav Müller-Franzes[5,†,*], Lorena Escudero Sánchez[1,†], Nicholas Payne[1], Alexandra Athanasiou[2], Michael Kalogeropoulos[2], Aitor Lopez[3], Alfredo Miguel Soro Busto[3], Julia Camps Herrero[3], Nika Rasoolzadeh[4], Tianyu Zhang[4], Ritse Mann[4], Debora Jutz[5], Maike Bode[5], Christiane Kuhl[5], Yuan Gao[6], Wouter Veldhuis[6], Oliver Lester Saldanha[7], JieFu Zhu[7], Jakob Nikolas Kather[7,8], Daniel Truhn[5,†], Fiona J. Gilbert[1,†]

1 University of Cambridge, Cambridge, United Kingdom
2 MITERA Hospital, Athens, Greece
3 Ribera Salud Group, Valencia, Spain
4 Radboud University Medical Center, Nijmegen, Netherlands
5 University Hospital RWTH Aachen, Aachen, Germany
6 University Medical Center Utrecht, Utrecht, Netherlands
7 University Hospital Carl Gustav Carus, Dresden, Germany
8 EKFZ Technical University Dresden, Dresden, Germany
† contributed equally
* corresponding author

## Abstract

Early detection of breast cancer is critical for improving patient outcomes. While mammography remains the primary screening modality, magnetic resonance imaging (MRI) is increasingly recommended as a supplemental tool for women with dense breast tissue and those at elevated risk. However, the acquisition and interpretation of multiparametric breast MRI are time-consuming and require specialized expertise, limiting scalability in clinical practice. Artificial intelligence (AI) methods have shown promise in supporting breast MRI interpretation, but their development is hindered by the limited availability of large, diverse, and publicly accessible datasets. To address this gap, we present a publicly available, multi-centre breast MRI dataset collected across six clinical institutions in five European countries. The dataset comprises 741 examinations from women undergoing screening or diagnostic breast MRI and includes malignant, benign, and non-lesion cases. Data were acquired using heterogeneous scanners, field strengths, and acquisition protocols, reflecting real-world clinical variability. In addition, we report baseline benchmark experiments using a transformer-based model to illustrate potential use cases of the dataset and to provide reference performance for future methodological comparisons.

# Introduction

Early detection of breast cancer is critical to improving patient outcomes and remains a global challenge, with millions of women diagnosed each year[1,2]. Mammography, often complemented by ultrasound, is the primary imaging modality for breast cancer screening. However, recent guidelines from the European Society of Breast Imaging (EUSOBI) recommended MRI as a supplementary screening tool for women with dense breast tissue[3].

Despite its high sensitivity, breast MRI presents practical challenges. Analyzing multiparametric MRI examinations is more time-intensive than reading mammograms and requires expert radiological interpretation. Therefore, implementing breast MRI, especially for population screening at scale, poses substantial challenges. This has prompted increasing interest in using artificial intelligence (AI), particularly deep learning methods, to assist in the interpretation and classification of breast MRI scans, thereby potentially improving diagnostic efficiency and enabling earlier detection of small breast cancers[4].

A major barrier to developing robust and generalisable AI models is the limited availability of large, diverse, publicly accessible breast MRI datasets. While several public datasets exist, they are typically derived from single-centre studies and are often carefully curated and homogenised prior to release[5–8]. As a result, these datasets may not adequately represent the heterogeneity encountered in real-world clinical settings. Furthermore, most publicly available datasets contain only cancer cases, which limits their utility for training classification models capable of distinguishing between malignant, benign, and non-cancerous cases.

To address these limitations, we present a large, multi-centre breast MRI dataset collected from six clinical centres in five European countries - Germany, the United Kingdom, Greece, Spain, and the Netherlands. This dataset encompasses various scanner manufacturers, clinical settings, and acquisition protocols, capturing the variability inherent to real-world clinical practice. It includes not only confirmed malignant lesions but also benign and control/unremarkable cases, making it the largest and most diverse publicly available dataset of its kind.

This dataset is part of the broader Open Consortium for Decentralized Medical Artificial Intelligence (ODELIA) project, a European Horizon initiative to develop privacy-preserving, AI-based diagnostic tools through swarm learning[9]. ODELIA brings together 12 partners from 8 countries, including academic institutions, research centres, and healthcare providers, to create robust, generalisable AI models for breast cancer detection using MRI. The dataset presented here represents a subset of the ODELIA consortium. It is intended to support the development, benchmarking, and validation of AI algorithms capable of functioning across diverse clinical environments.

# Methods

In the following section, we outline the data acquisition, labeling, and image processing related to the dataset presented.

## Previous Datasets

A summary of publicly available breast cancer MRI datasets is presented in **Table 1**. Except for a single dataset, all originate from a single institution. This lack of institutional diversity limits the heterogeneity typically encountered in real-world clinical settings, thereby hindering the ability to assess and improve the generalisability of AI algorithms. Furthermore, most existing datasets are designed primarily for tumour segmentation tasks and consist exclusively of malignant (cancer) cases. This restricts their utility for developing diagnostic models to classify scans as malignant, benign, or non-cancerous. To the best of our knowledge, the dataset presented in this work is the largest publicly available breast MRI dataset to date that includes malignant, benign, and non-cancer cases, collected across multiple centres.

| | **Region** | **Only Cancer** | **Annotation Level** | **Studies [N]** | **Centres [N]** |
|---|---|---|---|---|---|
| **Advanced-MRI-Breast-Lesions**[10] | Unknown | Unknown | Pixel* | 632 | 1 |
| **DUKE-Breast-Cancer-MRI**[11] | USA | Yes | Pixel | 922 | 1 |
| **MAMA-MIA**[5] | USA | Yes | Pixel | 1506 | >18** |
| **BREAST-DM**[7] | China | No | Pixel | 232 | 1 |
| **FastMRI Breast**[8] | USA | No | Case | 300 | 1 |
| **ODELIA** | EU | No | Case | 741 | 6*** |

**Table 1: Other public datasets for breast cancer MRI studies,** *only for 200 studies, ** ensemble of 4 datasets with an unknown number of centres involved, ***7 datasets

## Data Acquisition

Breast MRI examinations were collected from the following six European medical centres between December 2006 and May 2024:

- CAM: Cambridge University Hospitals, Cambridge, UK
- MHA: Mitera Hospital, Athens, Greece
- RSH: Ribera Hospital, Valencia, Spain
- RUMC: Radboud University Medical Center, Nijmegen, Netherlands
- UKA: University Hospital Aachen, Aachen, Germany
- UMCU: University Medical Center Utrecht, Utrecht, Netherlands

Each centre contributed between 31 and 250 studies. The CAM centre contributed two datasets, one for screening and one for symptomatic cancer patients. Imaging protocols adhered to institution-specific clinical standards, ensuring diverse data representation. Details of the protocols and scanner hardware can be found in the Acquisition Technique section of the **Supplementary Materials**. Ethics approval was obtained at each centre individually, and studies were conducted in accordance with local regulations. Detailed ethics approval numbers are provided in the Ethics Declaration section.

## Data Annotation

Expert radiologists at each centre classified lesions in both the left and right breast based on histopathological or 2-year follow-up information (ground truth). Initially, lesions were categorized as:

- **No lesion:** No contrast-enhancing lesion is visible
- **Benign lesion:** A contrast-enhancing lesion is visible but confirmed benign via biopsy or two-year follow-up.
- **Malignant lesion (DCIS):** A lesion of type ductal carcinoma in situ (DCIS)
- **Malignant lesion (Invasive)**: A malignant invasive lesion
- **Malignant lesion (Unknown)**: A malignant lesion of unknown specific type

## Image Processing

First, MRI scans were converted from DICOM to NIfTI, and a standardized naming scheme was assigned. To achieve this, a script was developed to extract metadata from each DICOM file. Using pattern matching and the series description, the files were categorized into dynamic T1-weighted (T1w) and T2-weighted (T2w) sequences. The T1w sequence was subdivided based on timing information, distinguishing pre-contrast images from the first to the n-th post-contrast images.

To enhance lesion visibility, a subtraction image was computed by subtracting the pre-contrast image from the first post-contrast image. We refer to this as the **"original" configuration**, as it represents the original image data in the local centres' Picture Archiving and Communication System, but stored in a standardized naming scheme. An example for UKA in the original image configuration is given in **Figure 1,** and for other centres in **Figure S1** of the **Supplementary Materials**.

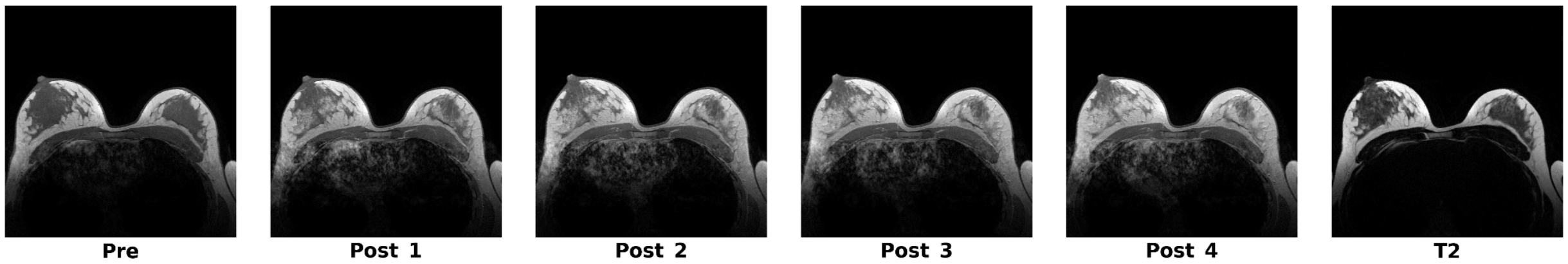

**Figure 1:** Axial slice of the pre-contrast (Pre), first to fourth post-contrast T1w sequences (Post), and the T2w sequence (T2) from the UKA “original” dataset.

For many machine learning applications, a more unified image data format is required. For this purpose, the T1w sequence was resampled to a resolution of 0.7 x 0.7 x 3.0 mm, and the T2w sequence was resampled to the T1w sequence. Subsequently, all examinations were separated into left and right breast regions by dividing the images at the centre. Furthermore, using a threshold to separate background and foreground, breasts were centre cropped or padded to 256 x 256 x 32 voxels. We refer to this as the **“unilateral” configuration**. An example of UKA in the unilateral image configuration can be found in **Figure 2**.

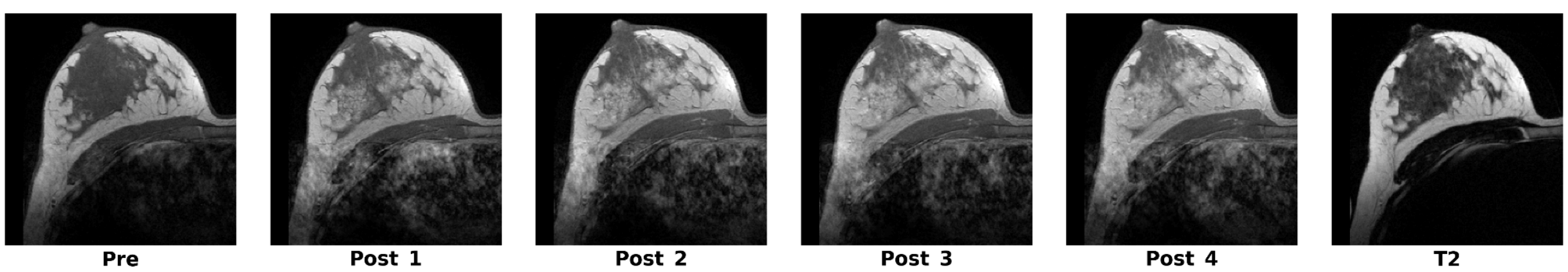

**Figure 2:** Axial slice of the pre-contrast, first to fourth post-contrast T1w sequences, and the T2w sequence from the UKA “unilateral” dataset.

## Annotation Processing

Due to the limited sample sizes of the malignant subtypes - namely DCIS, unknown, and invasive - all malignant classes were aggregated into a single class for subsequent analysis. For consistency, in cases of multiple lesions, only the most severe lesion was taken, prioritising malignancy over benign findings. Labels were mapped to codes as follows:

- No Lesion: 0
- Benign Lesion: 1
- Malignant Lesion: 2

# Data Records

The dataset is hosted on Hugging Face and can be accessed at https://huggingface.co/datasets/ODELIA-AI/ODELIA-Challenge-2025 . It includes breast MRI images and corresponding metadata for lesion classification and data splits.

## Files and Formats

Each institution is assigned a separate subfolder, named according to its initials. Within each institutional folder, data are organized into two subdirectories:

- “data” – containing the MRI images
- “metadata” – containing annotation and split configuration files

The original and unilateral dataset configurations follow the same folder structure.

**Image Files**

All MRI images are stored in 16-bit unsigned integer NIfTI format (.nii.gz). The dataset includes dynamic T1w sequences and T2w sequences, with the following naming convention:

- Pre.nii.gz – Pre-contrast T1w image
- Post_1.nii.gz to Post_n.nii.gz – First to n-th post-contrast T1w images
- Sub_1.nii.gz – Subtraction image (computed as the difference between the first post-contrast and pre-contrast images)
- T2.nii.gz – T2-weighted sequence

**Annotation File**

Lesion annotations are provided in “annotation.csv”. In the unilateral version, the columns “Lesion_Left” and “Lesion_Right” are merged into a single “Lesion” column. The annotation file contains the following fields:

- UID – Unique identifier for each study, matching the image folder names
- PatientID – Unique patient identifier
- Age – Patient’s age at the time of examination (in days)
- Lesion_Left – Lesion classification code in the left breast
- Lesion_Right – Lesion classification code in the right breast

**Split File**

To facilitate model evaluation, each institutional dataset is divided into five stratified cross-validation folds, ensuring no patient overlap across splits. The “split.csv” file contains:

- UID – Unique identifier for each institution
- Split – Indicates whether the sample belongs to the train, validation (val), or test sets
- Fold – Fold index for cross-validation (default: 0)

# Data Statistics

The dataset comprises 741 breast MRI examinations from 741 women, collected from six European medical centres. The mean age of participants was 54 ± 11 years, with examinations conducted between December 2006 and May 2024.

291 women exhibited no lesions, 146 women were diagnosed with benign but no malignant lesions, and 304 women had malignant lesions (**Figure 3a**). CAM contributed most patients with and without malignant lesions, while UKA contributed most patients with benign lesions (**Figure 3b**). In total, the dataset includes 978 breasts without lesions, 195 with benign lesions, and 309 with a malignant lesion (**Figure 4a**). CAM contributed most breasts with and without malignant lesions, while UKA contributed most breasts with benign lesions (**Figure 4b**).

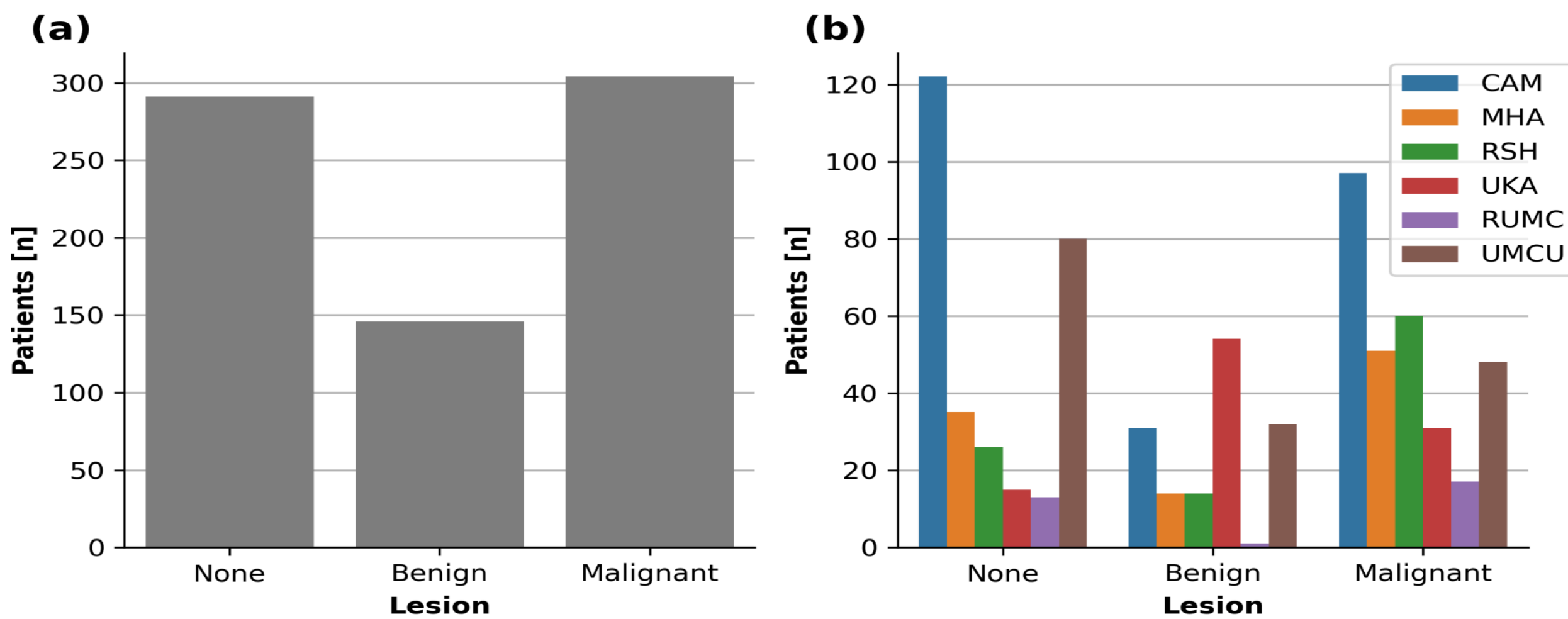

**Figure 3: Lesion distribution among patients.** (a) Overall lesion counts. (b) Lesion counts stratified by medical centres.

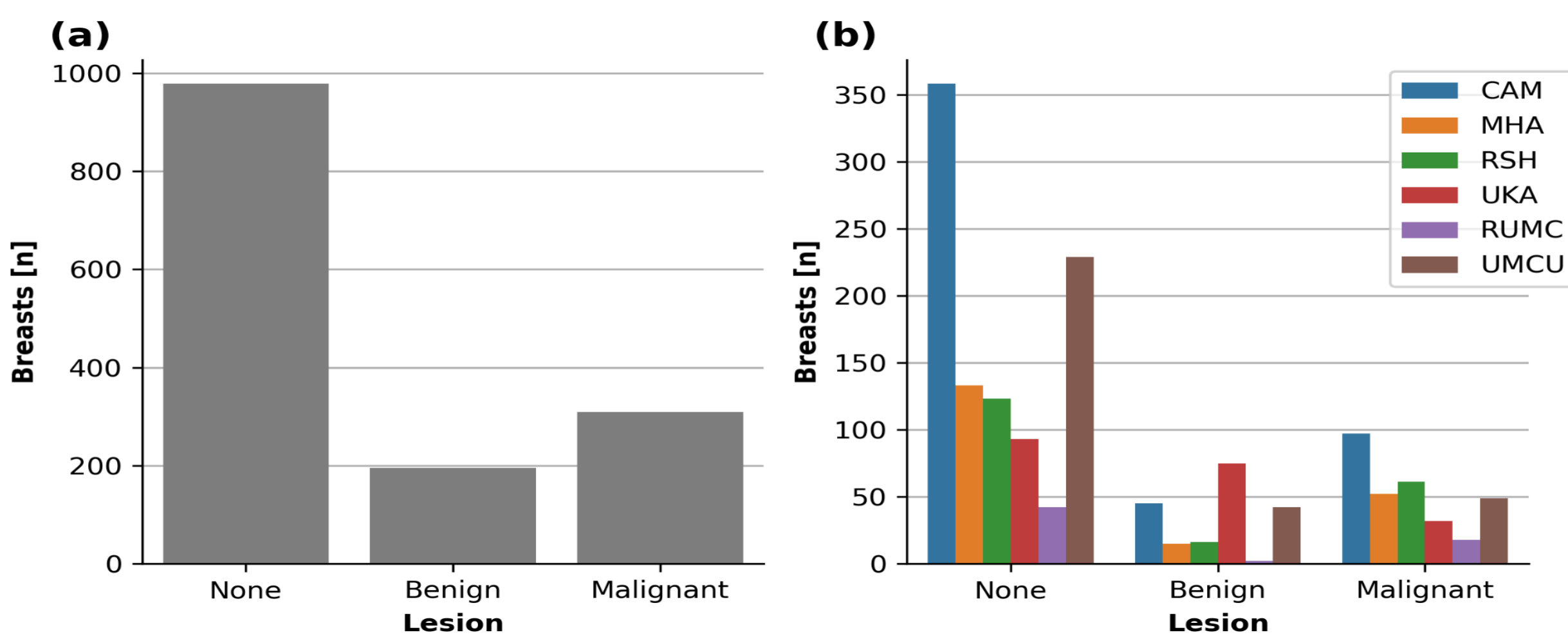

**Figure 4: Lesion distribution in the left and right breasts.** (a) Overall lesion counts. (b) Lesion counts stratified by medical centres.

# Technical Validation

## Study Design

To establish a benchmark and reference for automated lesion classification in breast MRI, we conducted two main experiments. In the first experiment, we trained a model using data from all centres except RSH and subsequently evaluated it on the test split from the same centres. We refer to this as the In-Distribution evaluation, as both the training and test samples were collected from multiple, but overlapping, centres. In the second experiment, we assessed the generalisation capability of the previously trained model by testing it on the independent RSH dataset. This evaluation is called Out-of-Distribution, as the test data originates from a centre unseen during training. For all experiments, we utilized the preprocessed, unilateral configuration of the dataset.

## Data Split

To ensure a robust evaluation, the multi-centre dataset was divided into three subsets: training, validation, and test sets. The training set comprised 64% of the data, the validation set 16%, and the test set 20%. A representative dataset was held out and exclusively used as an independent test set. We applied five-fold cross-validation, whereby each sample was used once as part of the test set.

## Model Architecture

We employed the Medical Slice Transformer (MST), a model specifically designed for volumetric medical imaging analysis[12]. MST leverages DINOv3[13] to extract a feature vector per slice and utilizes attention-based transformer mechanisms to aggregate all slices into a single prediction for the entire volume. The model received the pre-contrast image, the first post-contrast subtraction image, and the T2-weighted image as inputs.

Training was performed using the AdamW optimizer with a learning rate of 1e-5. The model was trained for approximately one hour on a single NVIDIA L40S GPU, with early stopping implemented to prevent overfitting. A batch size of 8 was used, and cross-entropy loss was employed as the objective function.

To enhance model robustness and generalisation, data augmentation techniques were applied, including random rotation, horizontal and vertical flipping, Gaussian noise injection, and cropping a window of 224×224×32 with random margins.

## Statistical Analysis

We evaluated model performance in distinguishing between no lesions, benign lesions, and malignant lesions using the area under the receiver operating characteristic curve (AUC), sensitivity, and specificity. Each label was treated as an independent binary classification task (i.e., Lesion vs. No Lesion, Benign vs. Others, and Malignant vs. Others). Sensitivity was calculated at a fixed specificity of 90%, and specificity was calculated at a fixed sensitivity of 90%. Additionally, we computed the AUC for each label and averaged the AUC across the three labels (macro AUC). Standard deviations were estimated using bootstrapping with 1,000 resamples.

## Results

A summary of the model's performance in classifying breast lesions is presented in **Table 2**. The receiver operating characteristic (ROC) curve and confusion matrix for the In-Distribution evaluation are shown in **Figure 5**, while those for the Out-of-Distribution test set are depicted in **Figure 6**.

**Table 2: Model's lesion classification performance.**

| | **Macro AUC** | **Micro AUC** | **Sensitivity*** | **Specificity*** |
|---|---|---|---|---|
| Fold 1 | 81.6 ± 2.3 | 85.9 ± 1.8 | 57.3 ± 5.6 | 61.5 ± 3.7 |
| Fold 2 | 81.5 ± 2.4 | 87.4 ± 1.9 | 66.4 ± 4.7 | 65.8 ± 5.9 |
| Fold 3 | 82.2 ± 2.7 | 91.3 ± 1.5 | 80.1 ± 4.2 | 75.2 ± 6.9 |
| Fold 4 | 75.0 ± 3.2 | 86.2 ± 2.0 | 64.1 ± 6.8 | 62.1 ± 5.3 |
| Fold 5 | 79.8 ± 2.6 | 88.5 ± 1.7 | 68.5 ± 5.5 | 60.8 ± 6.1 |
| Overall | 79.0 ± 1.3 | 87.4 ± 0.8 | 68.4 ± 2.4 | 63.3 ± 3.2 |
| Out-of-Distribution | 63.1 ± 3.5 | 78.3 ± 2.8 | 57.5 ± 7.3 | 26.7 ± 6.9 |

*Sensitivity at 90% Specificity and vice versa. All values are expressed as a percentage.

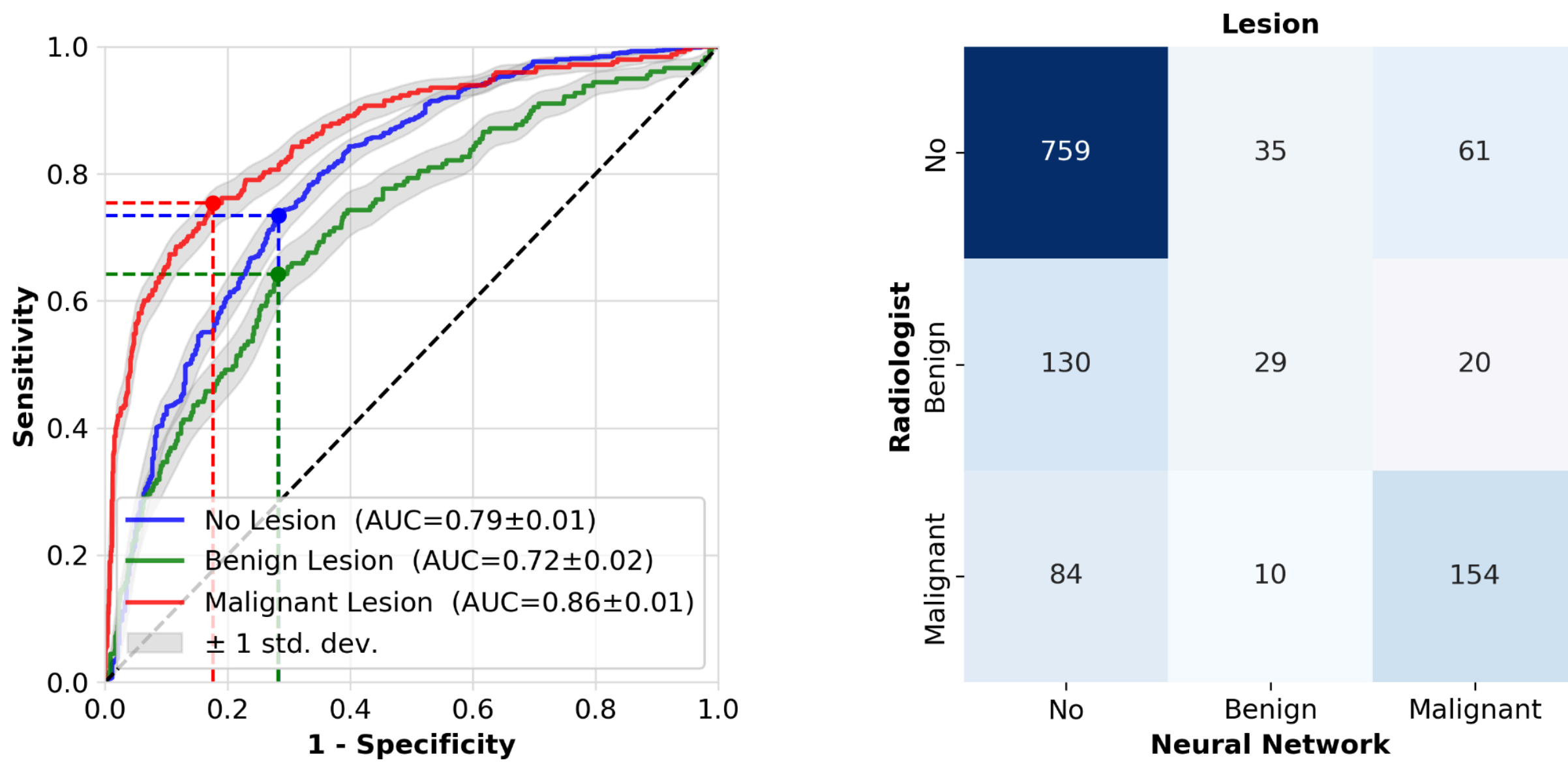

**Figure 5: Model's lesion classification performance for the In-Distribution test set.** (a) Receiver Operating Characteristic (ROC) curves depicting the sensitivity and specificity of the neural network in classifying lesions as "No Lesion" (blue), "Benign Lesion" (green), and "Malignant Lesion" (red). (b) Confusion matrix comparing the neural network's classifications to the radiologist's assessment.

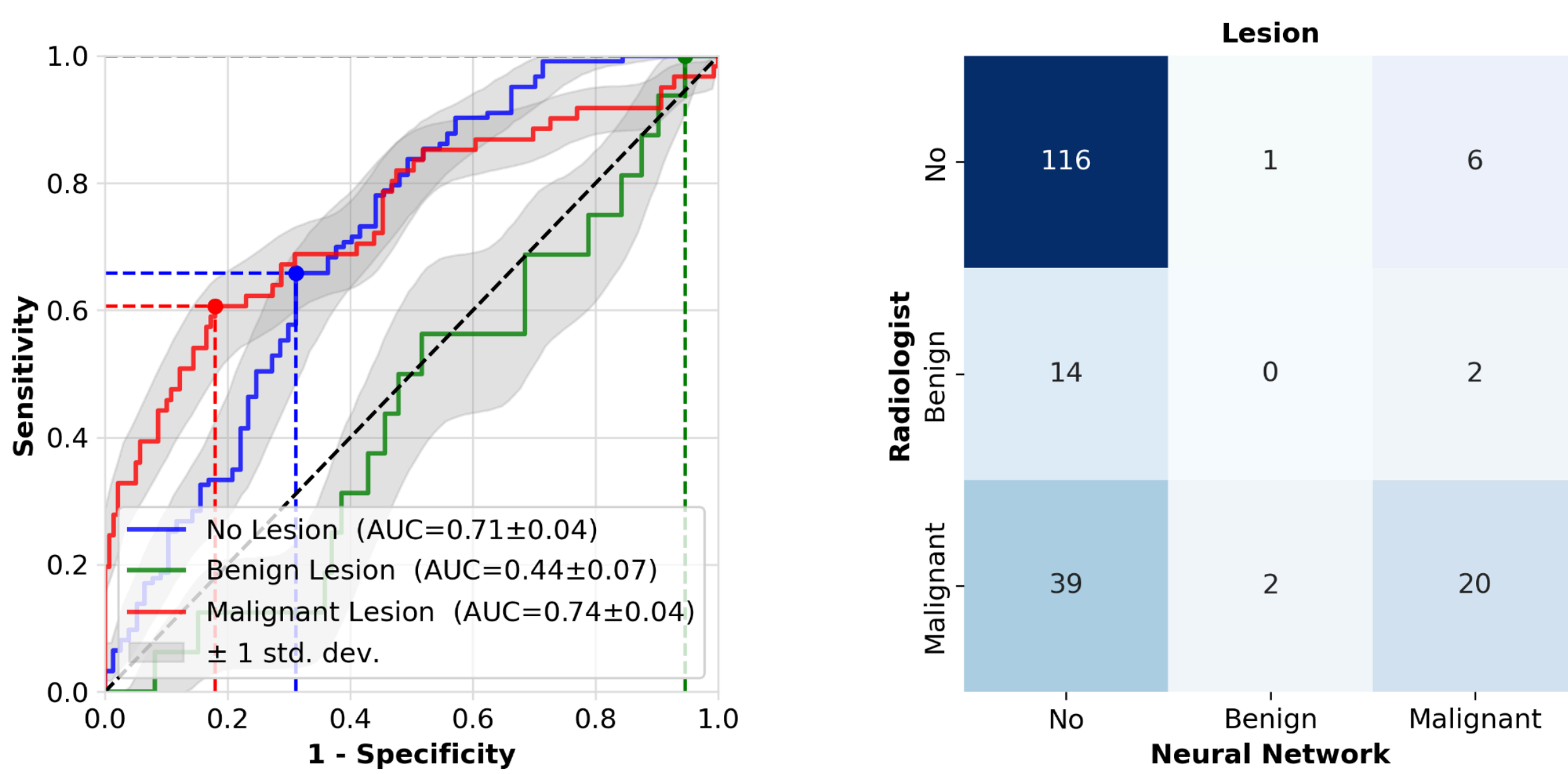

**Figure 6: Model's lesion classification performance for the Out-of-Distribution test set.** (a) Receiver Operating Characteristic (ROC) curves depicting the sensitivity and specificity of the neural network in classifying lesions as "No Lesion" (blue), "Benign Lesion" (green), and "Malignant Lesion" (red). (b) Confusion matrix comparing the neural network's classifications to the radiologist's assessment.

# Usage Notes

This dataset is available under the Creative Commons Attribution-NonCommercial 4.0 International (CC BY-NC 4.0) license. This license permits use, distribution, and adaptation of the dataset for non-commercial purposes, provided that appropriate credit is given.
Users must cite this publication when utilizing the dataset and acknowledge its source. By accessing or using the dataset, users agree to comply with the CC BY-NC 4.0 license terms, including proper attribution and non-commercial usage.

# Data Availability

The datasets generated and validated in the present study are publicly available in the Hugging Face Repository at: https://huggingface.co/datasets/ODELIA-AI/ODELIA-Challenge-2025

# Code Availability

The code is publicly available at: https://github.com/mueller-franzes/odelia_breast_mri
The model weights are available at https://huggingface.co/ODELIA-AI/MST

# Acknowledgements

- Part of the data used in this publication was managed using the research data management platform Coscine with storage space granted by the Research Data Storage (RDS) of the DFG and Ministry of Culture and Science of the State of North Rhine-Westphalia (DFG: INST222/1261-1 and MWK: 214-4.06.05.08 - 139057).
- Hugging Face for hosting the dataset.
- Meta AI for providing code and weights of the DINO models

# Funding

- The ODELIA project has received funding from the European Union's Horizon Europe research and innovation programme under grant agreement No 101057091.

# Competing Interests

- D.T. received honoraria for lectures by Bayer, GE, and Philips and holds shares in StratifAI GmbH, Germany and in Synagen GmbH, Germany.
- J.N.K. declares consulting services for Bioptimus, France; Panakeia, UK; AstraZeneca, UK; and MultiplexDx, Slovakia. Furthermore, he holds shares in StratifAI, Germany, Synagen, Germany, and Ignition Lab, Germany; has received an institutional research grant by GSK; and has received honoraria by AstraZeneca, Bayer, Daiichi Sankyo, Eisai, Janssen, Merck, MSD, BMS, Roche, Pfizer, and Fresenius.
- F.J.G., N.P. receive research support from GE Healthcare and Bayer.
- All other authors declare no competing interests.

# Ethics declarations

- CAM: 1) IRAS Project ID 122652, 2) IRAS Project ID 251317, 3) IRAS Project ID 143891, 4) IRAS Project ID 260281
- MHA: local approval received by letter on January 30th, 2023
- RSH: Local ID 23/430-E
- RUMC: Local ID 2024-17192
- UKA: Local ID EK 24-087 (23-006)
- UMCU: Because the dataset was fully anonymized, the local Ethics Board waived the requirement for ethical review on 5 March 2024.

# Supplementary Materials

## Imaging Protocols

**Table S1: Acquisition Hardware**

| | CAM (BRAID1) | CAM (TRICKS) | MHA | RSH | RUMC | UKA | UMCU |
|---|---|---|---|---|---|---|---|
| **Manufacturer** | GE | GE | Siemens | Philips | Siemens | Philips | Philips |
| **Scanner** | SIGNA Artist | DISCOVERY MR750 | MAGNETOM, Vida | Achieva | Skyra, TrioTim, Prisma_fit, Avanto | Achieva, Ingenia | Achieva, Ingenia |
| **Field Strength [T]** | 1.5 | 3 | 3 | 1.5 | 1.5 and 3 | 1.5 | 1.5 and 3.0 |
| **Coil** | 8-channel breast coil | 16-channel breast coil | 18-channel bilateral breast coil with frontal, circumferential, and axillary elements | Double Breast seven element surface coil (Invivo Corporation) with immobilization paddles | Varies from 4, 16, to 18 channels: 4ch like the one from UKA and 18ch like the one from MHA. No padding. | Double breast four-element surface coil (Invivo) with immobilization paddles | 7-channel dedicated bilateral breast coil |

**Table S2: Acquisition Parameters for the Dynamic T1w-sequence**

| | CAM (BRAID1) | CAM (TRICKS) | MHA | RSH | RUMC | UKA | UMCU |
|---|---|---|---|---|---|---|---|
| **Sequence** | GR | SPGR | GR | GR | GR | GR | GR |
| **3D** | Yes | Yes | Yes | yes | yes | No | yes |
| **Fat Suppression** | Yes | Yes | Yes | No | No | No | yes |
| **Echo Time [ms]** | 3.4 | 3.8 | 1.7 | 3.2±0.1 | 2.0±0.3 | 4.6±0.1 | 2.1±0.4 |
| **Repetition Time [ms]** | 6.9±0.1 | 7.1 | 4.8 | 5.8 | 5.5±0.2 | 262.7±19.8 | 4.4±0.8 |
| **Flip Angle [°]** | 10 | 12 | 10 | 18 | 16.0±2.0 | 90 | 9.8±0.6 |
| **Slices** | 68 to 116 | 112 | 104 to 122 | 110 to 130 | 144 to 176 | 25 to 31 | 106 to 222 |
| **Slice Thickness [mm]** | 2 | 2.8 | 2 | 2 | 1 | 3.1±0.2 | 1.9±0.1 |
| **Acquisition Matrix** | 512 | 512 | 360 to 456<br>256 to 256 | 400 to 448 | 416 to 448 | 512 to 560 | 352 to 672 |
| **Field of View [mm]** | 350 | 350 to 380 | 309 to 392<br>201 to 250 | 310 to 387 | 360 to 360 | 280 to 400 | 339 to 427 |
| **Post-contrast Acquisitions [N] *** | 2 | 4 to 5 | 4* | 5 | 4 to 5 | 4 to 5 | 7 |
| **Acquisition Timing [s]** | 65 seconds | 9.4 seconds* Total of 48 post-contrast phases (view sharing) Every 9th given, so effectively 84 seconds | 114 seconds x 1 average = 1:16 min, each dynamic x 5 | 83 + 41 for UF sequence, all following 83 | 71.67 to 205.42 | 70 + 20 and all following 70 | 63s for each dynamic phase |
| **Contrast Agent** | Gadobutrol (0.1 mmol/kg body weight) | Gadovist (0.1 mmol/kg body weight) | Gadobutrol (0.1 mmol/kg body weight) | Gadoteric acid (Clariscan) 0.1 mmol/kg body weight = 0.2 mL/kg body weight | Dotarem, gadavist (0.1 mmol/kg) | Gadobutrol (0.1 mmol/kg body weight) | Gadobutrol (0.1 mmol/kg body weight) |
| **Injection Rate** | 3 mL/s | 3 mL/s | Regular: 3 mL/s<br>If vein is poor: 2 mL/s | 3 mL/s, if vein is poor 2 mL/s | 3 mL/s | 3 mL/s | 1 mL/s |
| **Clearing** | 25 mL Saline flush | 25 mL Saline flush | 30 mL Saline Flush | 30 mL Saline Flush | 30 mL Saline Flush | 30 mL Saline Flush | 30 mL Saline Flush |

**Table Legend:** Mean and standard deviation provided when values differed. *Except one case with only 3. Abbreviations: GE = Gradient Echo, SPGR = Spoiled Gradient Echo

**Table S3: Acquisition Parameters for the T2w-sequence**.

| | CAM (BRAID1) | CAM (TRICKS) | MHA | RSH | RUMC | UKA | UMCU |
|---|---|---|---|---|---|---|---|
| **Sequence** | SE | SE | SE | SE | SE | SE | SE |
| **3D** | No | No | No | Yes | No | No | No |
| **Fat Suppression** | No | No | No | No | No | No | Yes |
| **Echo Time [ms]** | 88±1 | 81±1 | 87±4 | 365±5 | 107±38 | 110 | 87±19 |
| **Repetition Time [ms]** | 5511±1184 | 4939±1216 | 3602±194 | 2000 | 4285±733 | 3980±185 | 5544±616 |
| **Flip Angle [°]** | 160 | 111 | 120 | 90 | 94±19 | 90 | 90 |
| **Slices** | 68 to 115 | 22 to 98 | 36 to 44 | 200 to 250 | 60 | 33 to 39 | 31 to 100 |
| **Slice Thickness [mm]** | 2 | 3.7±0.8 | 4 | 1.5 | 2.5 | 3.1±0.2 | 2.2±0.5 |
| **Acquisition Matrix** | 512 | 512 | 416 to 832 | 432 to 528 | 320 to 384 | 512 to 672 | 256 to 560 |
| **Field of View [mm]** | 350 | 320 to 380 | 336 to 405 | 300 to 387 | 340 to 360 | 280 to 400 | 320 to 429 |

**Table Legend:** Mean and standard deviation provided when values differed. Abbreviations: SE = Spin Echo

# Examples

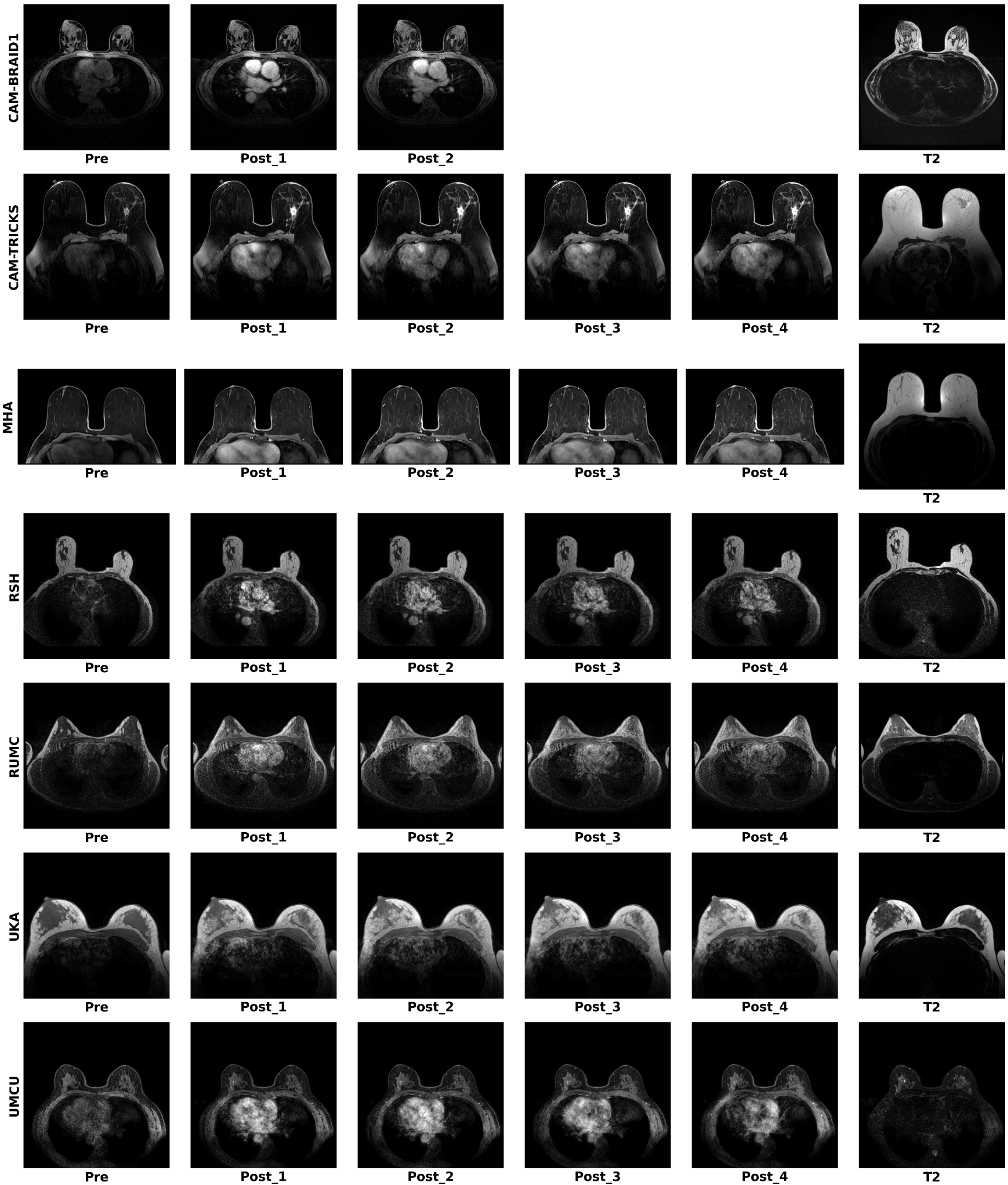

**Figure S1:** Axial slice of the pre-contrast, first to fourth post-contrast T1w sequences, and the T2w sequence.